**eXtended Reality for Autism Interventions: The importance of Mediation and Sensory-Based Approaches**

Valentin Bauer[1], Tifanie Bouchara[2], Patrick Bourdot[1]

[1] LISN, Université Paris-Saclay, Orsay, France
[2] CNAM-CEDRIC, HéSam Université, Paris, France

**Abstract**

eXtended Reality (XR) autism research, ranging from Augmented Reality to Virtual Reality, focuses on socio-emotional abilities and high-functioning autism. However common autism interventions address the entire spectrum over social, sensory and mediation issues. To bridge the gap between autism research and real interventions, we compared existing literature on XR and autism with stakeholders' needs obtained by interviewing 34 skateholders, mainly practitioners. It allow us first to suggest XR use cases that could better support practitioners' interventions, and second to derive design guidelines accordingly. Findings demonstrate that collaborative XR sensory-based and mediation approaches would benefit the entire spectrum, and encourage to consider the overall intervention context when designing XR protocols.

**Corresponding author**
Valentin Bauer, PhD student
LISN, VENISE team,
Université Paris Saclay,
Orsay, France
valentin.bauer@limsi.fr

# I. Introduction

Autism Spectrum Disorder (ASD) is a pervasive neurodevelopmental disorder with a worldwide prevalence of around one percent (Lord et al. 2020), and three main features (American Psychiatric Association 2013): communication and interaction disorders, focused interests, and sensory modulation disorders. In particular, sensory disorders affects around 90% of autistic people over the different sensory channels and largely contribute to the atypical child development (Robertson and Baron-Cohen 2017; Schaaf et al. 2011). In particular, they lead to specific features: a tendency to focus on the details rather than on the overall appreciation of a situation (Frith 1989), a perception that the world is going too fast (Gepner, 2018), various executive disorders, e.g. resistance to change (Lord et al. 2020), an "associative" way of thinking, meaning that autistic people need to gather lots of similar experiences and then compare it to make sense of a situation (Grandin 2009), or co-occurring conditions, e.g. attention-deficit (Lord et al. 2020). Since autistic people belong to a spectrum, they display various sensorimotor and cognitive abilities, some requiring substantial support, being non-verbal and with cognitive disabilities, and others requiring "minimal support to complete academic work". (Bottema-Beutel et al. 2021).

While early, structured and individualized interventions are advised to address autism challenges, multiple approaches exist (Sandbank et al. 2020). Naturalistic Developmental Behavioural Interventions (NDBI) draw upon behavioural theories of learning to "teach skills chosen from a developmental sequence in naturalistic environments" while "using natural rewards" (Sandbank et al. 2020; Schreibman et al. 2015). They include the Early Start Denver Model (ESDM) (Rogers et al. 2012), the Applied-Behavioural Analysis (ABA) (Lovaas 1987), and communication strategies, e.g. the Picture Exchange Communication System (PECS) (Flippin et al. 2010), or Makaton (Montoya and Bodart 2009). After that, the Training and Education of Autistic and other Communication Handicapped Children (TEACCH) (Mesibov et al. 2004) focuses on a "structured environmental design and self-monitoring" (Sandbank et al. 2020). Then, psychodynamic approaches concentrate on the relationship between the self and the others, as well as the interaction of forces within the person (Midgley et al. 2021). Then, sensory-based approaches use varied activities to enhance the integration of multisensory interactive processes by the body in order to improve cognitive abilities (Cascio et al. 2016). They include the Sensory Integration Therapy (Ayres 1972; Schoen et al. 2018) or Snoezelen (Lancioni et al. 2002). At last, integrative approaches use elements from different approaches that best suit the patient's needs (Klein and Kemper 2016). Yet, interventions often present challenges due to their low practical feasibility, e.g. lack of flexibility of the tools, hard-to-get training, or expensive cost (Lang et al. 2010). This paper suggests using technology-based approaches to support them.

Technology-based interventions using with video games on various mediums such as tablets, desktop computer, or robots, are promising to overcome these issues, because they are often appealing for autistic children (Grynszpan et al. 2014; Mazurek et al. 2015; Sandbank et al. 2020). In particular, recent reviews have highlighted that Virtual Reality (VR) (Bradley and Newbutt 2018; Dechsling et al. 2021; Mesa-Gresa et al. 2018; Parsons and Cobb 2011; Parsons et al. 2017) and Augmented Reality (AR) (Berenguer et al. 2020; Khowaja et al. 2020; Marto et al. 2019) are promising to complement common autism interventions, because they allow to create secure and individualized spaces, with precise control over the different stimuli, the rehearsal of exercises, and the recording of the user's actions. Moreover, they present a good acceptability and usability with Head-Mounted Displays (HMDs) (Malihi et al. 2020; Newbutt et al. 2020) and room-centric spaces, e.g. Cave Automatic Virtual Environments (CAVE) (Cruz-Neira et al. 1992; Mesa-Gresa et al. 2018). In this paper, VR and AR are referred under the term eXtended Reality (XR), because what belongs to the real and virtual worlds is a matter of proportions according to Migram's virtuality continuum (Milgram and Kishino 1994). Contrary to low-immersive devices (e.g. desktop computers), XR technologies offer embodied multisensory capabilities which are highly engaging for autistic individuals (Miller and Bugnariu 2016). This article extends the previous XR design surveys focusing on XR interventions for autistic users (Bozgeyikli et al, 2018), or NDBI through the use of VR (Dechsling et al., 2021), by adopting a participative design perspective and thus not limiting to a specific type of intervention. It compares a theoretical state of the art, through existing literature guidelines, with a state of therapeutic uses, practices and insights built from interviews with autism stakeholders, mainly including practitioners, which are together compared to create XR design guidelines for autism interventions close to stakeholders' needs.

So far, most XR studies focus on socio-emotional abilities, hardly considering the autistic sensorimotor and cognitive particularities (Bozgeyikli et al. 2018; Lorenzo et al. 2019; Spiel et al. 2019; Valencia et al. 2019). Hence, they exclude the lower part of the spectrum, having difficulties to work on such skills (Bozgeyikli et al. 2018; Parsons et al. 2019). Yet, Parsons et al. (2019)'s results from a two-year seminar including 240 autism stakeholders revealed that perceptual specificities have to be more considered during the design process so that to better support practitioners' interventions. To that respect, using participative design is advised to give a voice to all stakeholders (Fletcher-Watson et al. 2019; Frauenberger et al. 2011; Parsons et al. 2019; Porayska-Pomsta et al. 2012).

Moreover, most XR studies display inherent biases, due to possible misunderstandings or anxiety from autistic users, which prevent from their full validation (Bozgeyikli et al. 2018). They often come from a lack of communication between the different stakeholders involved, i.e. academics, practitioners, autistic individuals (Pellicano et al. 2013, 2014), a non-consideration of autistic perception, non-ecological tests, or non-collaborative designs (Bozgeyikli et al. 2018; Parsons et al. 2019). Yet, apart from the recent Dechsling et al. (2021)'s study which extends NDBI to XR, no XR design recommendations are rooted in common interventions. Hence, a new set of XR user-centered design guidelines has to be derived, based on autism stakeholders' advice, a knowledge of common autism interventions, and the existing XR autism literature. This paper aims at offering such guidelines.

The contributions of this paper are threefold. First, it presents and analyzes how autism interventions are conducted through semi-structured interviews with autism stakeholders, i.e. mainly practitioners, but also autistic people with or without their families, and academics. Second, the focus of XR studies is compared with stakeholders' needs to derive possible XR uses for autism interventions. Third, existing XR design recommendations are compared with stakeholders' insights to derive XR guidelines for autism interventions. After presenting the methodology, the results section is divided into three subsections following the three main contributions announced above. The last section discusses the findings and suggests future directions for the XR autism research field.

## II. Methods

This section details the methodologies used, first to study the existing XR literature for autism interventions, and second to conduct and analyze the semi-directed interviews with autism stakeholders.

### A. *Method for the article selection from the literature*

Our XR literature review focuses on sensorimotor disorders, mediation, and participative design, to represent all autism stakeholders' views. Online queries were made with Google Scholar engine, to use the "cited by" option for each paper and reference more recent studies with similar interests. They combined the keywords: "Autism" OR "ASD", "Virtual Reality" OR "Virtual Environment" OR "Augmented Reality" OR "Mixed Reality" OR "Technology" OR "Digital Tool", "Multisensory Environment" OR "Sensoriality" OR "Mediation", and "Participative Design". The inclusion criteria concerned autism-focused XR studies, and computer-related fields when relevant for the XR medium, i.e. robots, touch-screen devices, or room-centric setups. No strict exclusion criterion was used, but articles focusing on stereoscopic displays were preferred, e.g. CAVE and HMDs. In total, 37 articles were selected (Alcorn et al. 2014; Aruanno et al. 2018; Bartoli et al. 2014; Bozgeyikli et al. 2018; Brosnan et al. 2019; Brown et al. 2016; Carlier et al. 2020; Constantin et al. 2017; Dautenhahn 2000; Dechsling et al. 2021; Duris and Clément 2018; Garzotto et al. 2017; Garzotto and Gelsomini 2018; Halabi et al. 2017; Kerns et al. 2017; Koirala et al. 2019; Krishnappa Babu et al. 2018; Kuriakose and Lahiri 2017; Lorenzo et al. 2019; Malihi et al. 2020; Maskey et al. 2019; Newbutt et al. 2016; Newbutt 2013; Pares et al. 2005; Parsons et al. 2019; Parsons et al. 2017; Robins et al. 2004; Spiel et al. 2017, 2019; Tang et al. 2019; Tarantino et al. 2019; Tardif et al. 2017; Tsikinas and Xinogalos 2019; Virole 2014; Wallace et al. 2010, 2017; Whyte et al. 2015).

### B. *Method for the semi-directed interviews*

Semi-directed interviews were conducted with French autism stakeholders, mainly including practitioners. After presenting the methodology, data analyzes are exposed. Figure 1 summarizes the interview methodology.

*Interviews*

To get in contact with autism stakeholders, personalized emails were sent to three Autism Resource Centers, 49 healthcare structures specialized in ASD and neurodevelopmental disorders, three associations representing autistic individuals and relatives, and 66 practitioners. Most contacts were found in the TAMIS address list. Moreover, proceedings of recent conferences about ASD, sensoriality and technology were identified, using queries with the following keywords: "ASD" OR "autism", "digital" OR "technology", and "sensoriality". In Europe, sixteen conferences or workshops appeared, from 2012 to 2019, mainly in France, except for two in Spain and one in England. We examined every attendee's profile and took contact when their activity focused on ASD, technology, and sensoriality. Furthermore, participant was asked for other contacts at the end of each interview. Other experts were discovered through papers, blogs, and CRA webpages. Only French participants were contacted so that to avoid language misunderstanding with the main author. Moreover, most participants were in Ile de France Paris area, to facilitate further in-situ investigations, e.g. observations of medical practice.

5Formal interviews were conducted with 34 stakeholders, including practitioners (n=29), people from the autism community (n=4) and academics (n=1). Table 1 summarizes their profiles, which tend to be representative of the distribution of all autism fields [Table 1]. Practitioners and academics are highly experienced, often with more than ten years of practice (n=21). Practitioners display various backgrounds, i.e. liberal (n=5), non-liberal (n=10), or both (n=14). They mainly use integrative (n=15), over psychodynamic (n=6) and NDBI (n=5) approaches. Autistic individuals and their relatives favored NDBIs (n=4), and often the ABA method. Technological-based interventions were often reported (n=21), and all autistic participants already used digital tools (n=4). All participants agreed to be contacted again for further questioning or future testing.

*Protocol*

Interviews aimed at better understanding how autism interventions are conducted, and what are the possible needs and viewpoints of participants regarding the use of digital tools as a support for interventions, and particularly XR. They were based on a semi-directed questionnaire targeting practitioners that was built in three phases as suggested by Lallemand and Gronier (2016): Phase 1: Demographics (2 questions); Phase 2: Interview body (30mn, 9 questions); and Phase 3: End of the interview (5mn - 2 questions). All questions were rephrased depending on both the participant and interview context. Table 2 presents the core of the questions in column 2, with respect to the interview phases in column 1 [Table 2], and the keywords that were used by the researcher for each original question to ask additional questions in column 3. Question 9 (Q9) was added after the sixth interview, because all participants mentioned the use of mediation activities. With autistic participants, non-appropriate question was removed (Q3), while two questions were added: QA1 looked at their viewpoints regarding the healthcare interventions that had experienced; QA2 asked for details about their atypical sensory perception. With academic participants Q3 was also removed, and technology-focused questions were more explored (Q10, Q11).

Interviews with autism experts last between 34' and 94' with a mean duration of 59'. They include 19 women and 11 men, and were conducted using phone (n=18), visio-conference (n=7), or face-to-face (n=5). With autistic individuals and/or their families, they last between 43' and 70' with a mean duration of 51', and were conducted using phone (n=3) or face-to-face (n=1). The critical incident technique (Flanagan 1954) was used in order to elicit more responses, i.e. participants were always asked to give precise examples of the elements they were speaking about, for instance by describing how went the last time when this element occurred. Every participant signed a consent form prior interview, mentioning they were free to stop whenever they wanted. All interviews were audio recorded. All data collected were anonymized for transcription and analysis, using the "itw" keyword followed by the interview number. For security reasons, no online cloud system was used, and transcriptions were made manually. Repetitions, hesitations, prosody were not taken into account during the transcription process.

*Data analysis*

Data analysis mainly used a bottom-up approach with inductive coding following the *grounded theory* method (Charmaz, 2006). The emerging concepts were sometimes compared with already existing classifications from the XR literature to get refined. The Grounded theory consists in extracting meaningful phrasings from the interviews, and then classify them into concepts and categories. The technique starts from no preconceived concepts but progressively build and refine them through multiple iterations between the created concepts and data. This analysis process stops when the classification becomes stable. Since some concepts and categories are sometimes intertwined, some names can be found in multiple concepts and categories.

Four iterations were conducted by only one researcher at different times (i.e. offset of several weeks) to reduce possible biases. Since all interviews were conducted in French, the concepts and categories were first created in French, and then translated and refined in English. Only inductive coding was used to analyze autism interventions not using digital tools. About possible interventions using digital tools and XR mentioned by participants, concepts were regularly compared with existing classifications found in surveys of the field, i.e. Bozgeyikli et al. (2018), Lorenzo et al. (2019), Mesa-Gresa et al. (2018), Parsons et al. (2019), Spiel et al. (2019), and Whyte et al. (2015).

### III. Results

Stakeholders' insights about autism interventions without and with digital tools are first detailed, and respectively summarized in tables 3 and 4, from the most to the least reported category [Table 3][Table 4]. Then, existing XR uses and goals from autism research, mainly drawing from two recent reviews (Berenguer et al. 2020; Mesa-Gresa et al. 2018), are compared with stakeholders' needs, the findings being summarized in table 5 from the most to the least reported category. Finally, the comparison between XR design recommendations from the literature with stakeholders' insights is presented and summarized in table 6. Throughout the article, the number X of participants mentioning each concept is written inside parenthesis (n=X), next to a literature reference if also mentioning it. Participants' quotes were translated from French to English. The term "autistic child" is often employed to refer to autistic individuals, as participants mainly spoke about autistic children.

*A. Analysis of common autism interventions*

This subsection details common autism interventions without and with digital tools from the stakeholders' viewpoint. It does not focus on specific methods, except when mentioned by more than seven participants.

*Common non-digital autism interventions*

About the intervention context, Individualize the intervention program and Structure the intervention program largely appear (respectively n=33, n=27), since children are often ritualized and display strong interindividual differences. Individualization requires practitioners to be "creative", as a psychometrician says, in order to create tailored rewards (n=16) and games (n=10) **starting from the child's interests** (n=32), and depending on the intended therapeutic outcomes. For instance, a team supervisor says: "if a child loves the cartoon "Cars" then we'll use boxes with Cars!". It concerns long-term projects (n=13) and the care methods used (n=6), but also the sessions (n=14), as they depend on the child's state, e.g. tired, nervous. Though, individualizing the intervention (n=15) remains a challenge, being the most-reported sub-concept about difficulties in interventions, especially with non-verbal children who require substantial support. About structuration, on a local level, it consists in **offering predictability** (n=27) of time (n=21) and space (n=19), e.g. using routines (n=17). For instance, a speech therapist always "sets the phone in the drawer to avoid unexpected ringing". This aims at making the child feel secure and prevents from possible meltdowns due to being overwhelmed by too many stimuli. Though, activity-related difficulties can arise (n=14), for instance since providing structured spaces might be easier in institutions. On a global level, interventions have to be **structured over different periods** (n=11) which can last from days to months, depending on the child, and consist in first observing the child's interests through free play (n=8), and then use them to work while strengthening the child-practitioner relationship (n=7). Yet, practitioners warn about not over-using structuration (n=6), as it does not prepare for real life (n=4).

Regarding the intervention process, most stakeholders advise to Engage children in their intervention program (n=33), to maximize the intervention outcomes. This relies on various elements, such as **strengthening the therapeutic alliance** (n=30) (i.e. child-practitioner relationship), using **playful activities** (n=30), or using **behavioural methods** (n=26). In particular, the therapeutic alliance consists in building a secure relationship then allowing to create a **sense of agency** (n=16), being a prerequisite according to NDBIs to then be able to work on challenging activities. Engaging the child requires working on the content, through individualized playful activities (n=27), e.g. music-based (n=14), as well as the context, e.g. by adapting the practitioner's behaviour (n=19). For instance, an occupational therapist said: "I whisper, speak quiet, speak up… or speak with […] Makaton gestures". **Behavioural methods** (n=26) mainly include using rewards (n=19), gradually increasing challenges (n=18), **alternating activities** (n=15), or **using compensation strategies** (n=14), often sensory-based (n=7). Yet, building therapeutic alliance remains challenging (n=9), especially with children requiring substantial support (n=9).

Work on sensoriality in intervention (n=26) widely appears due to widespread sensory disorders, with two main concepts, i.e. **Work on Sensory Processing Disorders** (n=21) and **Provide multisensory environmental adaptations** (n=20). Practitioners often rehabilitate hypo/hyper sensitivities through sensory habituation (n=15). In particular, multisensory spaces (n=8) allow to "first remove distressing elements and then to increase them" (n=2). Other practices consist in modulating the audio-visual speed (n=7) or working on the multisensory binding (n=5). Sensory strategies aim at maximizing the attention and engagement. To that respect, various techniques are used, targeting the hyposensitive channels, for instance with sensory loads to regulate the child sensorimotor balance (n=13), or hypersensitive channels, e.g. with sensory protections (n=12), or make the space neutral (n=11).

At last, it is necessary to Evaluate Individual State over time (n=29), on the long (n=28) and short terms (n=21). On the long term, evaluations are often conducted at the intervention start, by combining standardized tests about sensorimotor (n=21), and psycho-developmental features (n=18), mainly based on questionnaires and interviews with relatives and the child (if possible), with direct clinical observations (respectively n=9, n=5). Combining evaluations at regular time intervals (n=8) allows to measure the child's progress. Then, on the short term, observations are used (n=18), by looking at specific individualized features, e.g. repetitive behaviours (n=2).

*Existing uses of digital tools in interventions*
Most participants use digital tools to support their interventions (n=27), since they are engaging, predictable (n=2), individualizable (n=2), repeatable (n=2), responsive (n=1), stable (n=1), can relieve from human interaction (n=2), and can allow to be wrong (n=1)(see Table 4). Participants mainly use tablets, being intuitive for autistic children (n=4), or desktop computers. Stakeholders' interests go beyond social skills, by including three main categories that are detailed below, i.e. Social skills (n=13), Assistive technologies (n=13), and Mediation/Well-being (n=10), as well as three minor categories, i.e. Education (n=8), Sensoriality (n=8), and Diagnosis (n=2). Various games are used since they have to be tailored to the child's abilities, the intervention context, and the intended therapeutic outcomes. Games with simple interfaces, little information and clear goals are preferred (n=5). To train Social skills, tablets or desktop computers are often used to display social scenarii (n=3) through solo (n=4) or group (n=2) activities. Then, Assistive technologies are highly individualized, as a team supervisor stresses when mentioning the case of a non-verbal autistic child, who knows all Disney cartoons by heart, and uses a tablet to play the sentences that she wants to say from them. About Mediation/Well-being purposes, any appealing game can be used (n=10), and a psychometrician notes that "video games allow to get into a prismatic relationship between the screen, the therapist and the child". Participants advocate for using digital tools in a controlled setting to avoid triggering new repetitive behaviours due to over-exposition to screens (n=5), as said by a speech-therapist citing Harlé and Desmurget (2012)'s work. This involves playing under the presence of an adult, with a time limit, and presenting video games as other activities, e.g. drawing. Yet, many practitioners remain hostile to digital tools, because consider that it sets a distance to the child (n=3), or due to a phantasm to be replaced by machines (n=1).

*B. Comparison between XR uses and stakeholders' needs*

The findings from the comparison between the XR autism research and stakeholders' insights are presented following the main categories that appeared: Mediation and Well-being (n=16), Social, Education and Cognitive Training (n=14), Sensoriality (n=13), and Assistive technologies (n=5) (see Table 5) [Table 5].

*Mediation and Well-being*

Whereas most stakeholders evoke the mediation and Well-being category (47%), it remains under-researched in the literature so far, and only includes three percent of the studies in Mesa-Gresa et al. (2018)'s review. Yet, Newbutt et al. (2020) previously raised this need, with 29 autistic children answering "It relaxes me and I feel calm" to the question "What could or would you use VR HMDs for?". Our participants report two main concepts, i.e. **Make child available for challenging tasks** (n=13), which aims at making the child in an optimal secure state before to perform challenging tasks, and **Support the therapeutic alliance through mediations activities** (n=6). Three use cases were proposed. First, scenarii inspired from multisensory spaces, often Snoezelen, were mentioned (n=8), considered as secure intermediate XR spaces allowing to strengthen the child-therapist relationship. Second, scenarii that are tailored to real-life stereotypic behaviours that children use to calm down, e.g. being in a car (n=1). A team director considered them as derivatives of stereotypies, that could "fill some sensory needs in a less stigmatizing way" and reach a "sensory saturation state" then allowing to perform challenging tasks. At last, creative activities appeared, drawing upon common interventions (n=3) such as music-making (n=2). Though, since some children may struggle to stop using these appealing XR simulations (n=3), participants suggest to gradually decrease any appealing stimuli before to stop (n=1) in order to prevent creating new focused interests.

*Social and Cognitive Training*

Whereas 41% of the stakeholders mention the Social and Cognitive Training category, it is the focus of more than 87% of the studies from the two literature reviews (Berenguer et al. 2020; Mesa-Gresa et al. 2018). Both the literature and stakeholders allude to real-life use cases, e.g. school playground. Two main concepts appear, i.e. **Train socio-emotional & Interaction abilities** (n=14) and **Train cognitive abilities** (n=6). The first concept consists in gradually habituating the child to daily situations through a secure space while benefiting from the XR embodied multisensory potential. To that respect, task-dependent features are modulated (e.g. number of people, or number of actions to perform) in various aversive situations (n=12) (e.g. medical examinations (n=6), school situations (n=4)) to gradually generalize the skills learned into the real life. Then, training cognitive abilities first include daily living skills (n=4), at home (e.g. brushing teeth) or outside (e.g. buying things in a supermarket). Other cognitive abilities are mentioned (n=3), such as attentional abilities through object discrimination tasks (Escobedo et al. 2014), by adapting NDBI methods (e.g. ABA) (n=1), or educational methods (e.g. Boehm-3) (n=1) (Boehm and Psychological Corporation 2000). Yet, such scenarii would only benefit to individuals requiring low support, being able to understand them. An autistic participant highlights another limit, i.e. "No manual could cover every possible social situation". Indeed, a psychometrician reports the case of an autistic adult who had learned the "right" way to say "hello" and got lost at his office due to the variety of situations that he encountered.

*Sensoriality*

Whereas participants highly mention the Sensoriality category (38%), including all psychometricians and occupational therapists, it remains under-explored in the literature. Indeed, only the *SEMI* project (Magrini et al. 2019) focuses on rehabilitating fine motor skills, with a kinect camera and four distinct applications. Three main concepts appear, i.e. **Rehabilitate sensory disorders** (n=10) and **Assess sensory disorders** (n=6), and **Work on action-reaction principles** (n=4). Two main use cases emerge: gradually add stimuli to reach the individual's tolerance sensory thresholds, and then conduct sensory habituation; start from a multisensory scene and gradually remove sensory information to assess the tolerance sensory thresholds. As a psychometrician says: "the challenge would be to recreate VR contexts allowing to assess the tolerance thresholds while conducting a therapy". Possible use cases draw from real-life scenarii, for instance a supermarket where every environmental information could be modulated (e.g. dimming the lights, adding noise, etc.), or multisensory spaces inspired from Snoezelen. This design choice mainly depends on the practitioner's preferences and the child's abilities. A psychologist notes that working on sensory disorders allows to include cognitive impaired people with non-verbalizable anxieties. At last, an occupational therapist warns about the impossibility to recreate in XR the richness of real-life sensory details.

*Assistive Technologies*

Whereas some participants mention the minor category Assistive technologies (15%), it remains absent from the literature. Two main concepts appear: **Provide context relevant-only information** (n=4) or **Support sensory strategies** (n=1). The first one can be achieved by adding and/or removing contextual information, e.g. adding individualized relaxing information on the walls of a real space (n=1), filtering or modifying colors (n=1), or withdrawing a background noise to enhance attention (n=1). Then, sensory strategies aim at offering a resourceful space to get the child in an optimal state then allowing to perform challenging tasks. To that respect, a participant suggested using a tipi-like space augmented by sounds and colors, inspired from the case of a child who could enter a tipi-like space in his classroom when feeling overwhelmed to get resourced and then get back to the class.

*C. Comparison between XR existing designs and stakeholders' design insights*

This section offers design guidelines from the comparison between the literature with the interviews 'findings (see table 6) [Table 6]. Its subsections follow the main categories and are organized following the main concepts.

*Task design*

**Individualization** is advised to cater for children and practitioners, regarding both the content (Bozgeyikli et al. 2018; Carlier et al. 2020; Dautenhahn 2000; Dechsling et al. 2021; Parsons et al. 2019; Parsons et al. 2017; Tang et al. 2019; Tarantino et al. 2019; Whyte et al. 2015); n=15), e.g. stimuli, tasks, and graphics, as well as the medium (Dechsling et al. 2021, n=1). For instance, a psychometrician mentioned using video games as "modelling dough", i.e. by adjusting every possible parameter. Whereas participants mainly suggest displaying familiar XR content (n=14) (e.g. drawings), the literature focuses on adapting the way the environment works (e.g. number of stimuli), sometimes with physiologically informed platforms (Krishnappa Babu et al. 2018; Kuriakose and Lahiri 2017).

As in common interventions**, engagement** aims at maximizing the intended intervention outcomes. According to the participants and the literature, it requires the environmental motor and cognitive complexity to be tailored to the child's abilities, the intervention context and the intended outcomes (Bozgeyikli et al. 2018; Carlier et al. 2020; Dechsling et al. 2021; Tang et al. 2019; Tarantino et al. 2019; Whyte et al. 2015, n=13). It also relies on prompting discovery by offering predictable and simplified audiovisual content (Bozgeyikli et al. 2018; Carlier et al. 2020; Dautenhahn 2000), including some unexpected events (Alcorn et al. 2014; Brown et al. 2016; Virole 2014, n=3), to create a "slight strangeness", as a psychologist said. Using common educational and NDBI principles is advised, i.e. rewards, imitation, and providing a sense of agency. Individualized rewards are often sensory-based (Bozgeyikli et al. 2018; Carlier et al. 2020, n=1) and are tailored to the performance (Constantin et al. 2017; Dechsling et al. 2021). Imitation consists in the practitioner imitating the child, or conversely, with promising outcomes about the therapeutic alliance and the training (Dechsling et al. 2021, n=1). Supporting the sense of agency consists in making the child active (Bozgeyikli et al. 2018; Dautenhahn 2000; Parsons et al. 2019; Spiel et al. 2019, n=14), and possibly including XR authoring activites (Bozgeyikli et al. 2018; Pares et al. 2005; Parsons et al. 2019; n=3), e.g. inspired from the painting VR game called *Tiltbrush* (Ying-Chun and Chwen-Liang 2018) (n=2). Whereas participants advise offering a high-level of control, e.g. choosing between different activites (n=6), the literature focuses on the actions during the gameplay. All these features, in addition to other strategies (Bozgeyikli et al. 2018; Kerns et al. 2017; Tang et al. 2019; Tsikinas and Xinogalos 2018; Whyte et al. 2015, n=5), such as using a scoring system (Bozgeyikli et al. 2018; Kerns et al. 2017), ensure that the child can have fun.

Considering autism **Sensoriality & Perception** allows to create well-suited designs for children from the entire spectrum. To that respect, structuring the time and space is advised (Bozgeyikli et al., 2018; Carlier et al., 2020; Dautenhahn, 2000 ; n=8), e.g. using time timers and repetition possibilities (Bozgeyikli et al., 2018; Carlier et al., 2020; Lorenzo et al., 2019 ; n=10). Due to the sensorimotor disorders, accessible interactions are preferred, such as touchless interaction (Bozgeyikli et al. 2018; Brown et al. 2016; Parsons et al. 2019, n=4), as well as offering varied interaction possibilities (Bozgeyikli et al. 2018; Pares et al. 2005; Parsons et al. 2019). To move around in space, embodied interaction is preferred over teleportation techniques (Bartoli et al. 2014; Bozgeyikli et al. 2016; Brown et al. 2016, n=1). Finally, making experiences meaningful can be achieved by drawing links with the real world, e.g. including familiar objects into XR (Bozgeyikli et al., 2018; Tang et al., 2019 ; n=1), and by considering autism perception during the design process, e.g. associative thinking (Dechsling et al. 2021; Virole 2014, n=2).

Since the child-practitioner relationship is at the core of interventions, including **collaboration** possibilities is fundamental. Indeed, a psychometrician said that the impossibility in many VR applications to be with the patient slows down the practitioners' acceptability of VR. Hence, the practitioner has to be able to prompt the child (Dechsling et al. 2021; Parsons et al. 2019, n=2), while the child should be able to see the practitioner in order to be reassured (Dautenhahn 2000, n=10). Moreover, controls can be shared between the child and practitioner (n=10), to build and strengthen the therapeutic alliance while making the child active. Only two participants note that the practitioner only has to be visible if context-relevant regarding the XR scenario, e.g. medical examinations.

*Protocol to conduct XR sessions*
Stakeholders insist on creating a secure **intervention context** to prevent biases, possibly due to participants' anxiety. Before and during sessions, it consists in offering predicability (Garzotto et al., 2017; n=5), making the experience meaningful (Dechsling et al. 2021, n=4), and supporting engagement (Bozgeyikli et al. 2018; n=8). On the long term, it relies on including the XR experiment as part of the overall intervention (Bozgeyikli et al. 2018; Robins et al. 2004, n=4) and/or planning a free-play period to get the child used to the system (n=3).

When the experience starts, **mixed methods** can help to assess the child's experience, by blending qualitative and quantitative methods that are adapted to their abilities, e.g. verbal or non-verbal. While no consensus exists so far regarding the methods used in XR studies, and a comprehensive overview is beyond the scope of this paper, our analysis made emerge major practices. Before sessions, clinical questionnaires assess developmental and sensorimotor abilities (Malihi et al. 2020; Maskey et al. 2019). They can be combined with a quantitative analysis of behavioural and physiological data before and during sessions to assess the child's state (Dechsling et al. 2021; Koirala et al. 2019; Kuriakose and Lahiri 2017). Custom questionnaires, often self-report, allow to infer the child's engagement (Aruanno et al. 2018; Garzotto et al. 2017; Garzotto and Gelsomini 2018; Tarantino et al. 2019), along with common XR questionnaires, e.g. targeting the feeling of presence (Wallace et al. 2010, 2017). Such evaluations have to consider practitioner's behaviours and their impact over the child (Dechsling et al., 2021). Their comparison at regular time intervals may yield insights about the overall child's progress. They also have to be combined with qualitative evaluations, i.e. observations (Brown et al. 2016; Pares et al. 2005; n=2), and interviews with practitioners, relatives, and the child if possible (Spiel et al. 2017; n=1). Yet, as a psychiatrist says, assessing the intervention outcomes is a huge challenge since "during the week the child does 300000 things".

*Design process*

Creating autism-friendly environments calls for using **participative design** (Bozgeyikli et al. 2018; Brosnan et al. 2019; Parsons et al. 2019; Spiel et al. 2019; n=5), to consider the needs from all autism stakeholders, including the individuals requiring substantial support (Bozgeyikli et al. 2018; Parsons et al. 2019; n=7). According to the participants and the literature, the **equipment** used has to depend on the overall healthcare context, for instance by being affordable (Parsons et al. 2019; n=2), but also the child's abilities, by being resistive and non-tethered (Bozgeyikli et al. 2018; Dautenhahn 2000; Newbutt et al. 2016); n=1), as well as the XR tasks (Dechsling et al. 2021), for instance by using AR to prompt the generalization of the skills learned. Whereas four participants are reluctant about using HMD with autistic children, mainly due to risks of isolation (n=2), the other participants advocate for a controlled use, namely, under the control of a practitioner. However, practitioners stress that the acceptability and usability may be child-dependent (n=9), and that wearing HMD may require using some sensory habituation beforehand (n=3). To prevent from inducing anxiety due to autistic children's ritualization, considering the **use context** is highly important by conducting experiments in clinical settings (Bozgeyikli et al. 2018; Parsons et al. 2019; n=1). Finally, whereas XR manufacturers recommend starting from the age of from 13 (Gent 2016), participants advocate for using task-dependent age recommendations, based upon the child's abilities and practitioner's expertise (n=7), e.g. 7/8 for sensory-based and relaxation purposes (n=2), 10/13 otherwise (n=5).

*Information presentation*

Displaying **little and clear information** is advised due to autism filtering difficulties, i.e. only task-relevant stimuli (Bozgeyikli et al. 2018; Carlier et al. 2020; Virole 2014), and audiovisual simplification (Bozgeyikli et al. 2018; Tarantino et al. 2019, n=4). Hence, to support understanding, adapting to the child's pace is recommended, for instance by adjusting the information speed (Tardif et al. 2017; n=13), or using minimal prosody (Carlier et al. 2020; Duris and Clément 2018; n=1). Indeed, as a psychometrician said, "learning processes require slowness".

The level of details and the realism of the graphics has to be **task-dependent** (n=7), e.g. realistic for social scenarii (n=5), and abstract and creative for mediation purposes (n=1). To that respect, a psychologist stressed that "realistic and non-realistic environments won't interest the same practitioners" (n=1). About social scenarii, using adjustable collaborative realistic settings are advised to train various skills, and especially turn-taking (Dechsling et al., 2021).

Since little information is advised, simplified cartoonish **avatars**, which include customization possibilities, are preferred to represent the others and the self (Bozgeyikli et al., 2018; Newbutt, 2013; n=1). The user avatar has to be positioned at real-world height (Bozgeyikli et al. 2018). Moreover, due to common perceptual filtering difficulties, only hearing others' avatars instead of both hearing and seeing them can be preferred (Newbutt, 2013).

## IV. Discussion

### A. Summary of Results

The first aim of this article was to check whether the existing XR research studies matched autism stakeholders' needs with respect to common interventions. Our comparison between the literature and 34 stakeholders' interviews, mainly including practitioners, revealed that whereas more than 87 percent of the studies focused on training socio-emotional abilities, the interviews highlighted to consider three main XR objectives, namely, well-being and mediation (47 percent), social and cognitive training (41 percent), and sensoriality (38 percent), as well as one minor objective called assistive technologies (5 percent) (see table 5). These objectives draw upon the main features of interventions without digital tools (see table 3), e.g. the use of mediation activities to build the therapeutic alliance, and the use of sensory-based approaches. Moreover, they are inspired from stakeholders' interests about technology-based interventions (see table 4). Indeed, the categories called social skills (38 percent) and education (24%) are close to the XR social and cognitive training category, and the categories called mediation and well-being (29 percent), and sensoriality (24%) have XR categories under the same name. Thus, research gaps emerged, which call for exploring more XR sensory-based and mediation approaches in order to better support the needs from all stakeholders, including individuals requiring high support. Furthermore, the main XR objectives reveal a strong inclusivity focus, to bridge the gap between a mere focus on training abilities and an only focus on changing the society to improve the well-being of autistic individuals. To that respect, our findings extend Parsons et al. (2019)'s research roadmap for the XR medium. The next subsection suggests possible use cases for implementing these categories, while stressing the benefit of using AR within a gradual VR-AR intervention.

The second aim of this article was to provide design guidelines that are representative from both the literature and the stakeholders. This paper revealed that whereas the recommendations presented similarities, differences emerged in terms of interests (see table 6). More specifically, stakeholders advocated for paying more attention to the overall intervention context in terms of the design process and of the XR protocols. Moreover, they stressed the importance of using collaborative designs. The third subsection will thus suggest XR autism design guidelines. In particular, methodological research directions will be outlined. At last, limitations to this work will be evoked.

### B. Suggestions of Use Cases for Future Autism XR Research

To train social abilities, using VR social scenarii (e.g. school playground) with a precise control of all environmental aspects by the practitioner is suggested. Yet, the autism literal way of learning calls for more research in order to maximize the generalization of skills learned from VR into real-life (Bozgeyikli et al. 2018). Moreover, whereas quickly shifting between different use cases could enhance the intended therapeutic outcomes (Dechsling et al. 2021), two participants stress that the sensory richness of real-life situations could not be recreated. Hence, VR training should be considered as part of the child's global intervention program, and within a gradual transition from VR to AR and finally the real life. For instance, a bakery scenario consisting in buying bread could first be trained in VR, to limit possible anxiety, and then in a real bakery with AR to withdraw distressing information (lights, etc.) while adding contextual elements (e.g. emotion detection). This VR-AR progression could help to gradually work in more ecological contexts due to AR (Berenguer et al. 2020). Even if more research is needed to enhance the generalization of the skills learned in AR (Berenguer et al. 2020), this VR-AR progression may present opportunities to maximize it by gradually confronting the child to real-life scenarii. This finding extends the complementary training VR and prosthetic AR roles proposed by Tarantino et al. (2019).

To make the child in an optimal secure state, three XR sensory-based approaches which draw upon common interventions emerged, i.e. multisensory relaxing spaces, derivatives of stereotypies, and mediation spaces.

Creating collaborative *multisensory relaxing XR spaces* largely emerged, often Snoezelen-inspired. A psychometrician described them as "sensory backpacks". Whereas VR use cases offer a precise control over all stimuli, VR displays risks of isolation (Parsons and Mitchell 2002). Furthermore, due to common autism proprioception and symbolization difficulties, children could misunderstand the representation of themselves and the practitioner as avatars. Thus, this could prevent the use of such scenarii for children requiring high support. While AR also presents risks of isolation (Berenguer et al. 2020), AR sensory approaches could overcome these issues, by perceiving the real surroundings and not using avatars. Thus, their low appearance in surveys (Berenguer et al. 2020; Mesa-Gresa et al. 2018) calls for more research. Possible use cases include real spaces where the proportion of real and virtual elements could be adapted to the session's needs, e.g. withdrawing posters on walls.

Some participants advise creating XR *derivatives of stereotypies* to replace with a non-stigmatizing XR approach the repetitive behaviours that children often use to calm themselves but are often considered as non-socially appropriate. Regarding that unexplored approach, both VR and AR call for more research. VR use cases could be considered if a whole context has to be recreated. For instance, one team supervisor mentioned the case of a boy who needs to be driven by his parents on Paris ring-road to get relaxed. Recreating this context in VR could support both the boy and his family. AR could be also used for recreating specific elements, e.g. spinning objects.

At last, XR *mediation activities* focus on strengthening the therapeutic alliance through collaborative free-play activities. Such "malleable" tasks, as a psychometrician said, could also prompt symbolization processes, as common mediation activities (Brun 2013). Yet, whereas the *activity theory* (Engeström et al. 1999) considers technology as mediating artefacts that encourage social processes, they remain unexplored in XR. Though, according to a psychologist, this approach could help to explore the concept of potential space (Winnicott, 1999), i.e. an intermediary area between the subjective experience and the objective reality for playful and creative experiences. In particular, such approaches could draw upon some non-XR digitally-augmented multisensory spaces which displayed promising outcomes regarding the therapeutic alliance (Brown et al. 2016; Garzotto and Gelsomini 2018; Gelsomini et al. 2019; Mora-Guiard et al. 2017; Pares et al. 2005; Ringland et al. 2014). They allowed to trigger multiple stimuli through body movements and manipulations with various interfaces (e.g. tangibles, reactive surfaces). *Mediate* was conducted in a large space (Pares et al. 2005), the *Magic Room* (Garzotto and Gelsomini 2018), *Magika* (Gelsomini et al. 2019), *Sensory Paint* (Ringland et al., 2014)*, and *Land of Fog* (Mora-Guiard et al. 2017) took place in smaller spaces, and the *Responsive Dome Environment* relied on a dome-like space where the child and practitioner sat together (Brown et al. 2016). Yet, these bespoke projects can be expensive and/or lack of flexibility. XR research should consider overcoming these limits with HMD-based AR.

XR sensory-based playful activities allow to work on sensory disorders to rehabilitate them, assess them, or work on action/reaction principles, often by adding/removing environmental stimuli. VR and AR could both provide solutions for creating such scenarii. While half of the participants mention VR Snoezelen-like scenarii, the other half mentions realistic situations ranging from VR to AR. In particular, a team director suggests starting in a fully controllable VR space, and then gradually going to AR. As previously mentioned about XR social scenarii, a possible use case could consist in recreating a real supermarket in VR, and then work in the real supermarket while using AR to remove distressing elements and add contextual help. Such AR setups could also be used as daily compensation strategies, as protection headphones commonly used in interventions. Hence, AR appears as highly inclusive, since it empowers all children to enter spaces usually considered as overwhelming. While such sensory-based AR scenarii remain under-explored, the current evolution of HMDs calls for more research in this area.

*C. Suggestions of XR Autism Design Guidelines*

XR task design draws upon many common intervention principles (see table 3), e.g. individualization, structuration, gradual challenges, or offering a sense of agency. Yet, contrary to the literature, stakeholders stress the need to create collaborative designs, as the success of common interventions largely relies on the therapeutic alliance. To that respect, as above-mentioned, AR seems promising, because it allows to connect with the familiar children's environment, by perceiving their usual practitioners and not using avatars. Then, two practitioners remarked that many design requirements that are listed in table 6 draw from common educational practices, and in particular the ones of Piaget et al. (1969) and Montessori and George (1964). As these practices often advise to use handling activities and AR can easily be combined with tangibles, AR seems well-suited to extend them.

Methodological insights regarding XR protocols expand previous recommendations (Bozgeyikli et al. 2018), by suggesting to focus both on the context of the intervention, and the evaluation of the outcomes. In particular, creating a secure space to then be able to conduct experiments was absent from the literature to our knowledge. On the short term it consists in offering predictability, i.e. before, during and after the sessions. To that respect, AR seems well-suited, due to the reasons above-mentioned. On the long term, XR protocols have to be divided into two main periods, i.e. first set the child in an optimal secure state, and then train specific abilities.

To assess the XR intervention outcomes, mixed methods have to be used, by blending qualitative and quantitative methods to adapt to the diversity of autistic profiles. While this evaluation remains a huge challenge due to the overall intervention program that children take part in, regular evaluations may provide insights about the progress made over time. So far, no standardized questionnaire exists to assess the autistic user's experience, and XR studies focus on different observable features. Yet, custom questionnaires have been created, for instance by Tarantino et al. (2019)'s who suggest to focus on four main features, i.e. the impact of photorealism, the understanding of real vs. virtual elements, the body movements, and the active exploration. Creating them involves not considering some commonly measured aspects with non-autistic individuals which are not relevant for autistic individuals. For instance, since assessing the feeling of presence, i.e. the feeling of "being here" (Biocca 1997), can already lead to ambiguous results with non-autistic children, it is even more questionable with some autistic children who may have difficulties to say if they feel present in real-life (Dautenhahn 2000). Moreover, whereas photorealism may engage non-autistic users, it produces the opposite effect with autistic users by distracting them (Tarantino et al., 2019). At last, since most studies use self-report questionnaires (Newbutt et al. 2020), they cannot be used with individuals who display intellectual impairments. More research is thereby needed to create such questionnaires, to be filled by practitioners with the help of children when possible, in line with Aruanno et al. (2018)'s study. They could also be combined with physiological data to get anxiety markers, e.g. heart rate (Kuriakose and Lahiri 2017). Yet, more research is needed to understand which relevant factors allow to get insights about complex perceptions (e.g. engagement), while using low intrusive technologies, such as light bracelets (Simões et al. 2018).

About the design process, XR studies have to use participative design, as suggested by prior research (Parsons et al. 2019), and to be conducted in clinical setups. This allows to make technological choices that are adapted to both the use context and the intended tasks. While the participants are positive about using HMDs, in line with the literature, they add that they have to be used in a controlled way, e.g. presence of a practitioner. This paper also revealed that participants would prefer to use task-depending age recommendations, which was absent from the literature to our knowledge, e.g. eight years old for mediation applications focusing on well-being purposes.

About information presentation, in addition to simplifying the audiovisual content, in line with the 2D games that are often appealing for autistic individuals (see table 4), and with Bozgeyikli et al. (2018)' findings, this paper revealed that the graphics realism has to depend both on the intended task and the practitioners' preferences. Indeed, practitioners being closer to psychodynamic or NDBI approaches may respectively prefer non-realistic creative scenarii, or realistic scenarii. Giving control over the audiovisual speed and prosody is also advised, as in Tardif et al. (2017)'s study. Indeed, decreasing the audiovisual speed may increase the child's understanding. At last, adding some unexpected events within a highly-structured XR space to enhance engagement connects with Remington et al. (2019)'s findings, which revealed that using too little or too many distractors hinders the attention of autistic children. This recommendation is also linked with compensation strategies that individuals with Attention-Deficit Hyperactivity Disorder (ADHD) often use to be able to focus, as well as with the Load theory (Lavie 2005), which states that neurotypical individuals can process and prioritize information until a certain cognitive and sensory threshold. Hence, future XR autism research should also consider individuals with neurodevelopmental disorders in general, due to its possible impact beyond the scope of autism,

*Limitations*

Findings must be considered in light of some limitations. Interviews were mainly conducted with practitioners, since XR guidelines aimed at supporting their interventions. Yet, this may have led to biases due to less considering the autistic individuals' viewpoints. Moreover, our study is limited by the fact that no single stakeholder could expertly provide insight about XR, but only suggestions based on their knowledge of common interventions with or without digital tools.

The evaluation of the included publications in terms of design features was conducted qualitatively and may contain inaccuracies. Also, no systematic review of the art was made, and the articles were mainly hand-searched.

The evolution of the understanding of the autism field by the first author throughout the interview process may have gradually changed his way to ask questions to the interviewees, and to adapt to them. This also may have led to elicit more answers over time. Furthermore, the first author' views may have influenced the results of the grounded theory process, but efforts were made to mitigate this bias by blending the inductive analysis process with a comparison from already existing classifications from the literature, as proposed by Charmaz (2006).

To complement and extend the findings from this paper, it will be particularly useful to conduct XR participative design workshops with autism stakeholders being representative from all autism fields.

## V. Compliance with Ethical Standards

Informed consent was obtained from all individual participants interviewed in this study.

**Figure Caption Sheet**

*Figure 1.* Method used to conduct and analyze the semi-directed interviews

# Figures

[Figure 1 top]

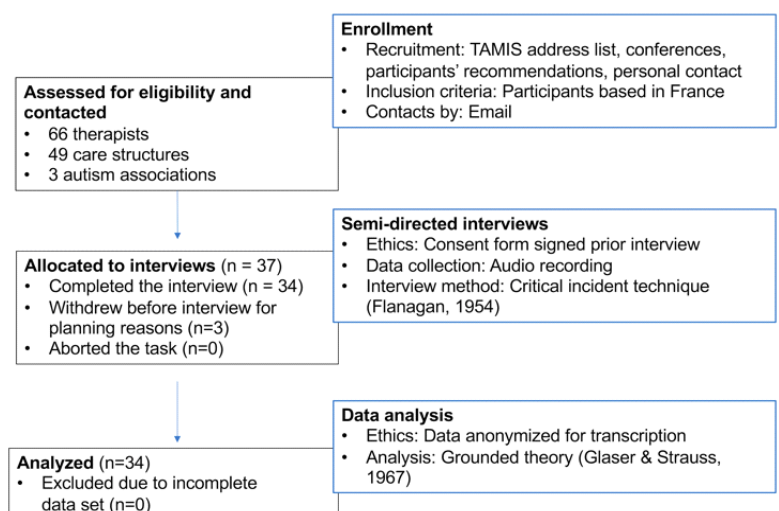

***Fig 1*** *Method used to conduct and analyze the semi-directed interviews*

# Tables

***Table 1*** *Profiles of the participants*

|  | **Practioners** | **Autistic community** | **Academics** |
|---|---|---|---|
| **Group size** | n=29 | n=4 | n=1 |
| **Sub-group characteristics** | - Psychologists (n=6)<br>- Speech-therapists (n=6)<br>- Team supervisors (n=5)<br>- Psychomotricians (n=5)<br>- Music-therapists (n=2)<br>- Psychiatrists (n=2)<br>- Occupational therapist (n=2)<br>- Specialized educator (n=1) | - High-functioning autistic individuals (n=2): alone or with family<br>- Mothers of autistic children requiring high support (n=2) | - Researcher in psychopathology and neurodevelopmental disorders (n=1) |
| **Gender (m/f)** | 10/18 | 3/1 | 0/1 |
| **Years of experience or age** | - 20 years and above (n=16)<br>- 10-20 years (n=5)<br>- 5-10 years (n=8)<br>- 0-5 years (n=1) | - 18-30 (n=1 - autistic individual)<br>- 30-50 (n=1 - autistic individual; n=2 – mothers of autistic children) | - 20 years and above (n=1) |
| **Rehabilitation trend** | - Integrative methods (n=15)<br>- Analytic approaches (n=6);<br>- NDBI approaches (n=5)<br>- Did not say (n=4) | - Preference for NDBI (n=4) | --- |
| **Field of activity** | - Liberal + institution (n=14)<br>- Institution (n=10)<br>- Liberal (n=5) | | --- |
| **Experience with digital tools** | - Computer / Tablet (n=21)<br>- Humanoïd robot (n=2)<br>- VR headset (n=1) | - Computer / Tablet (n=4)<br>- VR headset (n=1) | None |

*Table 2: Semi-directed questionnaire addressed to practitioners*

| | | Basis of question | Additional keywords/phrasings to get more precisions |
|---|---|---|---|
| **Phase 1: Demographics** | | Q1 – General approaches used in interventions | NDBI/psychodynamic/sensorimotor; Collaboration with other practitioners. |
| | | Q2 – Practitioner's background about autism interventions | Years of experience; Liberal or non-liberal activity. |
| **Phase 2: Core interview** | | Q3 – Typical autism intervention *(removed for non-practitioners)* | Examples of easy session; Examples of difficult session. |
| | | Q4 – Evaluation methods used before/during intervention | Sensory profile; Impact of sessions; Generalization of skills; Role of relatives. |
| | | Q5 – Individualization of intervention over time | Number and planning of sessions; Impact of atypical perception. |
| | | Q6 – Common intervention practices *(added after the 6th interview)* | Typical exercises; Difficulties experienced; Rewards; Individualization; Relationship with the practitioner; Structuration of time and space; Routines. |
| | | Q7 – Use and role of playful activities during sessions | Motivation; Engagement. |
| | | Q8 – Multisensory rehabilitation practices | Tools; Environments; Type of stimuli (visual, audio, tactile); Method used. |
| | | Q9 – Mediation activities in interventions | Music; Dance; Fine art; Relationship with the practitioner. |
| | | Q10 – Digital tools already used by the practitioner, and possible needs or viewpoints to that respect *(more explored with academics)* | Medium (i.e. tablets, robots, VAR); Game examples; Collaboration with the practitioner; Acceptability by children and practitioners; Ideal digital tool. |
| | | Q11 – Virtual or augmented reality environments that the practitioner would be interested in *(more explored with academics)* | Adaptability; Evaluation; Sensory processing disorders; Social aspects; Covering the entire autism spectrum. |
| | | QA1 - Viewpoints about current interventions *(added for autistic participants)* | |
| | | QA2 – Atypical autistic perception *(added for autistic participants)* | |
| **Phase 3: Closing up** | | Q12 – Any element that the practitioner wants to add | -- |
| | | Q13 – Other experts relevant to contact | -- |

*Table 3: Interventions evoked by the participants without the use of digital tools*

| Category | Concept | Sub Concept | Examples / Details |
|---|---|---|---|
| **Individualize interventions (n=33)** | Start from child's interests (n=32) | Create game depending on the purpose (n=10) | Game has to depend on interests (n=4), skills (n=2), fears (n=1) |
| | | Individualize rewards (n=16) | |
| | Individualize communication method (n=17) | Use alternative communication system (n=16) | Use pictograms / gestures, e.g. PECS (n=14) & Makaton (n=7) systems |
| | | Communicate through mediation activites (n=4) | Use any game (n=3) or object (n=1) |
| | Individualize sessions (n=14) | | Session has to depend on the child's state, e.g. fatigue (n=6) |
| | Build personalized project (n=13) | | Set personalized long-term goals based on evaluations (n=13) |
| | Individualize care method (n=6) | | No method suitable for every child exists (n=6) |
| **Engage children in their intervention program (n=33)** | Build and strengthen the therapeutic alliance (n=30) | Use individualized playful activites (n=27) | Use mediation activities (n=24), e.g. music (n=14), fine arts (n=5) |
| | | | Use any appealing games (n=12), e.g. video games (n=5) |
| | | | Use autistic interests: sensory stimulation (n=6), music (n=5), machines (n=4), geometric shapes (n=3), things that spin (n=3), bubbles (n=3) |
| | | Adapt the practitioner's behaviours depending on the child (n=19) | Be slow and enveloping (n=12), e.g. adjust voice intensity (n=4), Do not over anticipate (n=3), Adopt the child's pace (n=2) |
| | | | Provide meaningful explanations (n=8) |
| | | | Understand the "autistic world" (n=8), e.g. Allow stereotypies (n=1) |
| | | Create a secure relationship with the child (n=14) | Create a secure relationship (n=9); Enjoy the activity conducted (n=7) |
| | Use playful activites (n=26) | Adapt game based on the intervention | Goal can be social (n=13), cognitive (n=5), sensory (n=4) |
| | | goal (n=25) | |
| | | Entertain the child and widen his interests (n=4) | |
| | | Increase the child's autonomy (n=2) | |
| | Use developmental and behavioural methods (n=26) | Use rewards (n=19) | Use sensory-based rewards (n=9), the token economy method (n=5), or verbal congrats (n=4) |
| | | Gradually increase the challenge (n=18) | Gradually adjust sensory loads (n=13), helping strategies (n=3), number of elements (n=3), therapist place (n=1) |
| | | The child imitates the practitioner (n=7) | |
| | | Use physical prompts (n=4) | |
| | | Set things available to prompt demands (n=3) | |
| | Provide a sense of agency (n=16) | | Give choice over a few tasks (n=16), offer possibilities to move (n=1) |
| | Alternate activities (n=15) | Alternate types of activity (n=15) | Use challenging vs. appealing (n=13), known vs. unknown tasks (n=1) |
| | | | Often change activities (n=3) |
| | | Alternate ways to work on activities (n=3) | Task can be done sitting (n=3), moving (n=3), outside (n=1) |
| | Use compensation strategies to relax the child (n=14) | Make child available for challenging tasks (n=12) | Use sensory stimulation (n=7), relaxing activity (n=4), start from stereotypic behaviours (n=2) |
| | | Give daily strategies to prevent meltdowns (n=4) | Use derivatives of stereotypies (n=2) |
| **Evaluate individual state over time (n=29)** | Assess child state on the long-term (n=28) | Conduct sensory evaluation (n=21) | Conduct sensory profiles (n=20), clinical observations (n=9), sensorimotor evaluation (n=4) |
| | | Conduct psycho-developmental evaluation (n=18) | Assess developmental state and autism severity (n=15), language skills (n=8), conduct clinical observations (n=5) |
| | | Ask questions to parents (n=10) | Use discussions (n=7) and questionnaires (n=3) |
| | Assess the child evolution on the short term (n=21) | Carry out observation of sessions (n=18) | Take notes (n=3), observe number of shared attentions (n=3) and stereotypic behaviours (n=2), new interests (n=3) |
| | | Assess child at different intervention times (n=8) | Carry out specific assessments (n=8) |
| | | Discuss with people knowing the child (n=8) | Discuss with relatives (n=8), people at school (n=2) |
| **Structure the intervention program (n=27)** | Offer predictability of intervention program (n=27) | Offer time predicability (n=21) | Show visual planning at session start (n=6) and timers (n=4) |
| | | | Finish session with appealing activity (n=4) |
| | | Offer space predicability (n=19) | Use clear learning areas (n=6), clean space before child arrives (n=3) |
| | | | Show pictures of activities before to start (n=2) |
| | | Use routines (n=17) | Ritualize activities and their organization (n=14) |
| | | | Repeat activites (n=6) |
| | | Use same methods in all activities/spaces (n=9) | Provide parental advice (n=9) |
| | Structure intervention over different periods (n=11) | Observe individual to detect his interests (n=8) | |
| | | Build the relationship with the child (n=7) | |
| | Don't overuse structuration (n=6) | Prepare for real life unpredictability (n=4) | |
| | | Over structuration can prompt ritualization (n=3) | |
| **Face difficulties in interventions (n=27)** | Face difficulties working with autistic individuals and their families (n=21) | Find the suitable individualized intervention approach (n=15) | Some sensory channels remain hard to stimulate (n=7) |
| | | | Assess the impact of the intervention (n=6) |
| | | | Help without stigmatizing (n=4) |
| | | | Make activities meaningful (n=2) |
| | | Build the therapeutic alliance (n=9) | Understand the child's actions and feelings (n=6) |
| | | | Handle heteroaggressivity (n=3) |
| | | Reassure parents regarding possible fears (n=8) | Fears of the methods used (n=6) and institution (n=4) |
| | Face difficulties due to the healthcare system (n=20) | Face activity-related difficulties (n=14) | Health environment differs from daily life (n=9) |
| | | | Face difficulties linked with liberal/non liberal activity (n=7) |
| | | | Healthcare equipment is often expensive (n=3), e.g. Snoezelen |
| | | Face difficulties external to one's practice (n=12) | Sensoriality in interventions remains new (n=6) |
| | | | Face lack of specialized practitioners (n=6) |
| **Work on sensoriality in intervention (n=26)** | Work on sensory processing disorders (n=21) | Conduct gradual sensory habituation (n=15) | Gradually adjust specific sensory loads (n=13) |
| | | | Work in flexible multisensory environment (n=8) |
| | | Modulate audiovisual information speed (n=7) | Use Logiral (n=6) (Tardif et al. 2017) or Youtube (n=1) |
| | | Work on multisensory representations (n=5) | Use sensory lottos³ (n=5); Gradually adjust sensory density (n=1) |
| | | Use contrasted sensory elements (n=3) | Use environmental contrast in multisensory spaces (n=2) |
| | Provide multisensory environmental adaptations (n=20) | Offer sensory loads to regulate the child sensorimotor balance (n=13) | Give specific simulation based on child particularities (n=9) |
| | | | Work in adjustable multisensory environment (n=8) |
| | | Use sensory protections (n=12) | Protections can be audio (n=12) (headphones), tactile (n=2), (smooth ground cover), visual (n=1) (sunglasses) |
| | | Work in a neutral environment (n=11) | Limit environmental sensory information: visual |

| | | | distractors (n=6) (posters on walls…), Neutral colour of walls (n=3) |
|---|---|---|---|

Token economy method can be situated as part of ABA methods. It is uses systematic reinforcements of target behaviours, with "tokens" that can be exchanged for other rewards.

*Table 4: Fields of Interventions already using digital tools evoked by the participants*

| Field | Sub field | Content Description | Game Name (when mentioned), Medium | |
|---|---|---|---|---|
| | | | 2D graphics | 3D Graphics / other |
| **Social skills (38% / n=13)** | Social skills and emotions (n=4) | Video modeling (n=2) | | Youtube or other applications (C/P, 3D, n=2) |
| | | Social scenarii (n=3) | *Autimo* (P, n=1)(Auticiel, 2015) | *JeMiME* (C,3D,n=1)(Grossard et al. 2019) *JeStiMuLE* (C,3D,n=1)(Serret et al. 2014) |
| | Group activities (n=2) | Video game workshop (n=2) | *Degrees of Separation* (C/PS, n=1)(Moondrop, 2019) | *Human Full Flat* (C/PS,3D,n=1)(No Brakes Games, 2016) *Ico* (C/PS,3D,n=1)(Sony Interactive Entertainment, 2011) |
| | | Storytelling workshop (n=1) | NS | Research project (R, Other, n=1) (Duris and Clément 2018) |
| **Assistive technologies (38% / n=13)** | Alternate and Augmented Communication (n=12) | Tailored pictograms (n=12) | *NikiTalk* (P, n=3)(La Rocca, 2019) *Snap Core* (C/P, n=2)(Tobii Dynavox, 2021) *LetMeTalk* (P, n=1)(appNotize UG, 2014) *Dis-moi!* (P, n=1)(Caulavier, 2013) *Proloquo2Go* (P,n=1)(AssistiveWare, 2013) *Avaz* (P, n=1)(Avaz, Inc., 2020) *CommunicoTool* (P, n=1)(C..Texdev, 2016) | NS |
| | Daily planning (n=1) | Visual plannings (n=1) | -- | NS |
| **Mediation / Well-being (29% / n=10)** | Rewarding activity (n=10) | *Any appealing game* | Application allowing to burst balloons (P, n=1) | Youtube Channels (C/P, NS, n=1) |
| | Mediation (n=4) | *Any appealing game* | *Angry birds* (all,n=2)(Rovio,2009) *Théâtre de Minuit*(C,n=1)(Dada Média, 1999) *Bumpy* (Cons, n=1) (Loriciels, 1989) Various brick breaker games (T/C, n=1) Video game displaying an aquarium(C, n=1) *Tetris* (all, n=1)(Pajitnov, 1984) *My Talking Panda* (P, n=1)(Sofia_Soft, 2017) *MyPiano* (P, n=1)(Trajkovski Labs, 2000) *Real Drum* (P, n=1)(Kolb Sistemas, 2012) *Noogra Nuts* (C/P, n=1)(Bengigi, 2012) Games based on « Simon Says » (all, n=1) *Talking Ginger* (P, n=3)(Outfit7, 2012) | *GTA Vice City* (Cons, 3D, n=1)(Rockstar North, 2015) |
| **Education (24% / n=8)** | Cognitive remediation (n=8) | Programming workshop (n=2) | NS | *RobAutisme* (C/R, other, n=1)(Sakka et al. 2018) *Scratch* Programming Language (C, other, n=1)(Resnick, 2006) |
| | | Appealing games puzzle, memory games, etc. (n=7) | *Cognibulles* (C,2D,n=1)(Virole and Wierzbicki 2011) *BitsBoard* (P, 2D,n=1)(2021) | *School* (P, n=1, other)(LearnEnjoy, 2012) *Watch'n'Learn* (P, n=1, other)(Peters, 2018) |
| | Narrative understanding (n=1) | Storytelling workshops (n=1) | Research project (R, n=1) (Duris and Clément 2018) | NS |
| | Daily living skills (n=1) | Brushing the teeth (n=1), etc. | *Ben le Koala* (P, n=1)(Happy Moose Apps, 2017) | NS |
| **Sensoriality (24% / n=8)** | Multisensory Binding (n=6) | Slowing down videos (n=4) | NS | Youtube (Any, n=1, other) *Logiral* (C/P, n=3, other)(Tardif et al. 2017) |
| | | Auditory understanding (n=1) | NS | Application for doing listening lottos (NS, n=1, other) |
| | | Audio-Tactile tangible (n=2) | NS | Tangible allowing to play music (TI, n=1, other) Theremin instrument (TI, n=1, other) |
| | Psychomotor disorders (n=2) | Fine motor skills (n=1) | NS | -- |
| | | Gross motor skills (n=1) | NS | *Just dance* (Ki, n=1, 3D)(Ubisoft, 2009) Bowling application (Ki, n=1, 3D) |
| | Body consciousness (n=1) | Storytelling workshop (n=1) | NS | *RobAutisme* (C/R, n=1, other) (Sakka et al. 2018) |
| **Diagnosis (6% / n=2)** | ―― | Sensory profile (n=1) | NS | *SensoEval* and *SensoMott* (P, n=1, other)(Gorgy, 2012) |
| | | Language understanding | NS | -- |

| | | | (n=1) | | | |

C: Desktop Computer; P: Tablet and Phone ; R :Robot ; TI: Tangible Interface; PS :Projected Screen; Cons : Video Game Console ; Ki: Kinect; NS: Not-specified

*Listening lottos* are listening games where the child has to listen to the sound that are presented and then associate them with the corresponding images.

*Table5: Fields of interest of participants regarding XR use cases for autism interventions*

| Category | | | Concept | | | Objectives | | Examples | |
|---|---|---|---|---|---|---|---|---|---|
| Name | Weight | | Name | Weight | | Name | Participants | Literature (from MG and B reviews) | |
| | P | L | | P | L | | | | |
| **Mediation & Well-being** | 47% / n=16 | MG: 3% B: 0% | Make child available for challenging tasks | n=13 | MG: 0% B:0% | Use individualized multisensory space (n=8) | Snoezelen-inspired (n=10), tipi-like spaces (n=1) | -- | |
| | | | | | | Use individualized real stereotypies (n=3) | Breaking glasses (n=1), Being on a trampoline (n=1), in a car (n=1) | -- | |
| | | | | | | Use audio-only appealing environment (n=2) | | -- | |
| | | | Support the therapeutic alliance through mediations activities | n=6 | MG:0% B:0% | Do creative activities (n=3) | Painting (n=2), Making music (n=2) | -- | |
| | | | | | | Do any appealing activity (n=1) | Any activity for the child | -- | |
| | | | | | | Use Snoezelen-like space (n=1) | | -- | |
| | | | Enhance Well-being though Physical Activity | n=0 | MG:3% B:0% | Give motivation to do sport (n=0, MG:3%, B:0%) | | *Astrojumper* VR exergame-SSS (Finkelstein et al. 2013), | |
| **Social & cognitive training** | 41% / n=14 | MG:87% B:95% | Train socio-emotional & Interaction abilities | n=14 | MG:69% B:70% | Train Social & Interaction abilities (n=13, MG :45%, B:55%) | Real-life social scenarii (n=12): School (n=4), Cafeteria (n=2), transports (n=2) | Interact with a virtual dolphin to learn communication – VR – RS (Cai et al. 2013); Collaborative activity requiring perspective-taking – VR – C (Parsons 2015); Extending pictogram communication systems – AR – T/P (Taryadi and Kurniawan 2018); Virtual Agents & Structure turn-taking activity– VR – NM (Bernardini et al. 2013) | |
| | | | | | | Anticipate fearful situations (n=7, MG: 3%, B:0% ) | Medical examinations (n=6), Transports (n=2) | Real-life spaces (MG:3%): e.g. Individualized fearful scenes (e.g. dogs) –Blue Room VR Environment (RS)(Maskey 2019) | |
| | | | | | | Train emotions (n=1, MG:21%, B:20%) | *NM* | Social scenarii (home scene, school bus, school library, tuck shop, physical education class on the playground)- Half-CAVE-VR (Ip et al. 2018), Facial expressions & emotions using an augmented storybook– AR – T/P (Chen et al. 2016) | |
| | | | Train cognitive abilities | n=6 | MG:18% B:25% | Train cognitive abilities (n=3, MG:5%, B:15%) | Adapting VB-MAPP (n=1) and Boehm concepts to XR (n=1) | Attention (n=0, MG:5%): Object Discrimination using small games, T-AR (Escobedo et al. 2014), Grasp objects – VR - NM (Manju et al. 2018) | |
| | | | | | | Train daily living skills (n=4, MG:13%, B:5%) | Indoor environment: Having a shower (n=2), Brushing their teeth (n=2) | Indoor environment: Brushing teeth - marker-based AR picture prompt to trigger a video model clip of a student (Cihak et al. 2016) | |
| | | | | | | | Outdoor environment: *NM* | Outdoor environment: Real-life social scenarii (e.g. supermarket ) - HMD-VR (Adjorlu et al. 2017); | |
| | | | | | | Train abstract thinking (n=1, | Implicit humor in social situations | Pretend-Play - PP & AR - C (Bai et al. 2015) | |

| | | | | | | MG:0%, B:5%) | (n=1) | |
|---|---|---|---|---|---|---|---|---|
| **Sensoriality** | 38% / n=13 | MG: 0% B:5% | Rehabilitate sensory disorders | n=10 | MG:0% B:5% | Rehabilitate sensorimotor disorders (n=10, MG:0%, B:5%) | Sensory habituation to real-life scenarii (n=7): School (n=3), Supermarket (n=2), Daily situations (n=2) | |
| | | | | | | | Sensory integration in multisensory spaces (n=6): Snoezelen-like (n=6) | *SEMI* Project – Four interactive games to train motor skills - AR – Ki (Magrini et al. 2019) |
| | | | Assess sensory disorders | n=6 | MG:0% B:0% | Assess sensory disorders (n=4) | Real-life scenarii (n=2), e.g Supermarket (n=1) | -- |
| | | | | | | | Non-realistic scenarii (n=2), e.g. Snoezelen (n=2) | -- |
| | | | | | | Assess progress over time (n=2) | | -- |
| | | | Work on action-reaction principles | n=4 | MG:0% B:0% | Get inspiration from real-life scenarii (n=3) | | -- |
| **Assistive Technologies** | 15% / n=5 | MG: 0% B: 0% | Provide context relevant-only information | n=4 | MG:0% B:0% | Add assistive virtual information (n=3) | Adding information to daily social situations (n=1), e.g. about emotions (n=1) | -- |
| | | | | | | Filter non-relevant information (n=2) | Filtering distressing noises (n=1) | -- |
| | | | Support sensory strategies | n=1 | MG:0% B:0% | Offer a resourceful sensory space (n=1) | Creating an AR tipi-life space (n=1) | -- |

AR: Augmented Reality, P: Participants, L: Literature, Comparison with Literature reviews in VR from Mesa Gresa (MG) and AR from Berenguer (B), RS (Room-Size display), HMD: Head-Mounted Display, T/P: Tablet or mobile phone; C: Desktop Computer, SS: Stereoscopic Surround-Screen, SM: Smartglasses, PP: Physical Props, NM: not mentioned, Ki: Kinect.

*Table6: Comparison of XR guidelines coming from the literature with suggestions from the participants*

| Cat. | Concept | Sub-Concept | Literature elements | Participants |
|---|---|---|---|---|
| **Task design** | **Individualization** | Content | *Individualize content*: Vary the tasks, interactions, stimuli, graphics (Bozgeyikli et al. 2018; Carlier et al. 2020; Dautenhahn 2000; Dechsling et al. 2021; Parsons et al. 2019; Parsons et al. 2017; Tarantino et al. 2019; Whyte et al. 2015), add familiar objects into XR (Tang et al. 2019), or onto the real-environment using AR (Tarantino et al. 2019), use physiological data to tailor the content, e.g. gaze (Krishnappa Babu et al. 2018), heart rate, skin temperature (Kuriakose and Lahiri 2017) | *Individualize content* (n=15): Set a familiar environment with reassuring elements (n=14), adjust every sensory information (n=5), possibly switch on/off every parameter (n=2), set open/closed space depending on individual (n=1), individualize rewards (n=1), integrate individualized and alternative communication systems (n=1) |
| | | Medium | *Individualize medium according to the child's preferences* (Dechsling et al. 2021): Use CAVE if HMD is not tolerated, desktop computer /tablet if more feasible | *Individualize medium according to the child's preferences (n=1)*: Use tablet/desktop computer if HMD is not accepted |
| | **Engagement** | Fun | *Make it playful*: Use a scoring system (Bozgeyikli et al. 2018; Kerns et al. 2017), challenges and hidden elements (Tang et al. 2019; Tsikinas and Xinogalos 2019), storytelling and short/long term goals (Tang et al. 2019; Whyte et al. 2015), non-linear gameplay (Tang et al. 2019), digital companion (Tang et al. 2019), immediate feedback, customization of avatars | *Make it playful (n=5)*: Use individual's interests (n=5) (i.e. music (n=3), circular elements (n=2), play around presence/absence (n=2), play with their own shadow (n=1), visualize progress (n=3) (e.g. score system, collectables), use feedbacks and feedforwards (n=2), use rewards (n=2) |
| | | Discovery | *Prompt discovery*: Use various non-concurrent elements (Bozgeyikli et al. 2018), e.g. movement (Bozgeyikli et al. 2018; Dautenhahn 2000), shapes (Bozgeyikli et al. 2018), audiovisual stimuli (Carlier et al. 2020), 3D animations (Bozgeyikli et al. 2018), use unexpected elements (Alcorn et al. 2014; Brown et al. 2016; Virole 2014) | *Prompt discovery*: Use various non-concurrent elements, e.g. audiovisual stimuli surrounding the child (n=1), use unexpected elements (n=3) |
| | | | *Broaden the child's attention* (Dechsling et al. 2021): Use eye-tracking to detect fixations | |
| | | Body perception | *See oneself in XR* (Bozgeyikli et al. 2018) | *See oneself in XR (n=1)*: See one's shadow (n=1) |
| | | Environmental arrangement | *Gradually increase/decrease*: motor and cognitive complexity (Bozgeyikli et al. 2018; Carlier et al. 2020; Dechsling et al. 2021; Tang et al. 2019; Tarantino et al. 2019; Whyte et al. 2015), i.e. number of elements, e.g. crowdedness, stimuli, dynamism, (Tarantino et al. 2019) or types of elements/reactions, e.g. shapes, avatar's reactions, prompts (e.g. instructions, gestures) (Dechsling et al. 2021) | *Gradually increase/decrease (n=13)*: number and types of stimuli (n=8), realism level (for social scenarii) (n=4), predictability (n=3), environment neutrality (n=3), number of controllable elements (n=1), number of distractors (n=1), dynamism (n=1), prompts (n=1) |
| | | Rewards | *Individualize rewards* (Dechsling et al. 2021; Tang et al. 2019): use personal (Kientz et al. 2013; Whyte et al. 2015), e.g. sensory-based (Bozgeyikli et al. 2018; Carlier et al. 2020) or generic rewards (Constantin et al. 2017), Often assess new child's rewards (Dechsling et al. 2021) | *Individualize rewards (n=1)* |
| | | | *Adjust rewards:* Consider the child's performance and progression (Constantin et al. 2017; Dechsling et al. 2021), e.g. antecedents and behaviours, and the child's abilities (Dechsling et al. 2021) | |
| | | Sense of agency | *Make the child feel in control* (Bozgeyikli et al. 2018; Dautenhahn 2000; Parsons et al. 2019; Spiel et al. 2019): Use child-initiated episodes, e.g. from what they like, or by making them choose between different actions/activites/game pathways (Dechsling et al. 2021; Pares et al. 2005; Parsons et al. 2019; Tang et al. 2019; Whyte et al. 2015), Adjust the information pace (Tardif et al. 2017), Make the environment respond to various actions (gestures, voice…)(Pares et al. 2005) | *Make the child feel in control (n=14)*: Use child-initiated episodes, e.g. from what they like, or by making them choose between different activities (n=6), Giving them the possibility to freely move (n=2) |
| | | | *Allow the user to author the XR environment* (Parsons et al. 2019) | *Allow the user to author the XR environment (n=3)*: record and repeat sounds or videos (n=3) |
| | | Imitation | *Use modelling partners who simulate situations* (Dechsling et al. 2021), e.g. peers, avatars, | *Use modelling partners who simulate situations* (n=1) |
| | | | *Use avatars who imitate the child* (Dechsling et al. 2021), e.g. language, play and body movements | *Use avatars who imitate the child* (n=1) |
| | **Sensoriality & perception** | Structuration of time & space | *Offer predictability* (Bozgeyikli et al. 2018; Carlier et al. 2020; Dautenhahn 2000): Avoid loud sudden sounds (Bozgeyikli et al. 2018) | *Offer predictability (n=8)*: Use little distractors (n=6), Alert about their presence, e.g. using stimuli (n=2), visual activity timers (n=3) |
| | | | *Give repetition possibilities* (Bozgeyikli et al. 2018; Carlier et al. 2020; Lorenzo et al. 2019): e.g. Routine in training (Bozgeyikli et al. 2018; Carlier et al. 2020) | *Give repetition possibilities (n=10)* |
| | | Interaction types | *Use accessible interaction types* (Bozgeyikli et al. 2018; Brown et al. 2016; Parsons et al. 2019), e.g. speech recognition (Halabi et al. 2017), touchless interaction (Bartoli et al. 2014) | *Use accessible interaction types (n=4)*: ergonomic controllers (n=4), tangibles XR controllers (n=3) |
| | | | *Offer various ways to interact* (Bozgeyikli et al. 2018; Pares et al. 2005; Parsons et al. 2019) | |
| | | | *Use motion-based embodied interaction* (Bartoli et al. 2014; Brown et al. 2016; Dautenhahn 2000) | *Use motion-based embodied interaction (n=1)*: no teleportation (n=1) |
| | | Meaningful experiences | *Draw links with the real world* (Bozgeyikli et al. 2018; Tang et al. 2019) | *Draw links with the real world (n=1)* |
| | | | *Make experiences meaningful* (Dautenhahn 2000): consider autism perception, e.g. visual memory (Bozgeyikli et al. 2018), associative way of thinking (Dechsling et al. 2021; Virole 2014) | *Make experiences meaningful (n=1)*: Use individualized communication system of the child in XR (n=1) |
| | **Collaboration** | Prompting & Reassurance | *A practitioner/relative can prompt the child* (Dechsling et al. 2021; Parsons et al. 2019), e.g. instructions, gestures, physical prompts | *The practitioner/relative can prompt the child (n=2):* to help (n=2) |
| | | | *The child can see the practitioner*: for reassurance (Dautenhahn 2000) | *The child can see the* practitioner (n=10): for reassurance (n=2), if context-relevant (n=2) |
| | | Shared controls | | *Share controls between the child and practitioner* (n=10) |
| **Protocol to conduct XR sessions** | **Intervention context** | Secure environment | *Give predictability:* Practitioners may wear the HMD and invite the child to manipulate it before to wear it (Garzotto et al. 2017) | *Give predictability (n=5)*: Make the planning clear before to start (n=3), Use pictograms showing what the XR space will look like (n=1), Practitioners may wear the HMD before the child (n=1) |
| | | | *Make the experience meaningful for the child*: Combine different elements and strategies (Dechsling et al. 2021) | *Make the experience meaningful for the child:* Use understandable vocabulary to present the experience (n=4), Make sure everything is understood prior to start (n=2) |
| | | | | *The practitioner can control every XR parameter (n=9):* See what the child sees (n=2) |
| | | Organization of sessions | | *Get the child used to the system during first sessions (n=3)*: Possibly use sensory habituation to the HMD (n=3), Use free play to detect the child's preferences (n=1) |
| | | | *Keep the child engaged:* Make short sessions with breaks (Bozgeyikli et al. 2018) | *Keep child engaged (n=8)*: Alternate work and relaxation activities in XR (n=5), Alternate work in XR and in reality (n=2); Keep sessions short (n=1) (e.g. around 15mn) |
| | | | *Include XR experiment as part of the global intervention:* Make long-term studies (Bozgeyikli et al. 2018; Robins et al. 2004) | *Include XR experiment as part of the global intervention (n=4):* Have regular sessions every week (n=2); Use XR at particular moments of therapy (n=1); Make long-term studies (n=1) |
| | | | *Establish detailed procedures and provide training for the practitioners* (Dechsling et al. 2021) | |
| | **Mixed methods** | Quantitative Evaluation | *Assess the child's state before the experience* (Malihi et al. 2020; Maskey et al. 2019) | *Assess initial state of the child (n=1):* Conduct sensory profile |
| | | | *Assess the child' experience*: No consensus exists: Collect behavioural data (Dechsling et al. 2021), e.g. repetitive behaviours (Pares et al. 2005), interaction logs (Dechsling et al. 2021), eye tracking (Dechsling et al. 2021; Koirala et al. 2019), Collect physiological data (Kuriakose and Lahiri 2017), Create custom questionnaires (Aruanno et al. 2018; Garzotto et al. 2017; Garzotto and Gelsomini 2018; Tarantino et al. 2019), Use common XR presence or anxiety questionnaires (Malihi et al. 2020; Wallace et al. 2010, 2017), Use autism intervention questionnaires (Malihi 2019; Maskey 2019), etc. | *Assess the child' experience and performance (n=2)*: Collect physiological data for stress measurement, e.g. biosensors, pressure sensors (n=2), log observational data (n=2) |
| | | | *Assess practioners' actions* (Dechsling et al. 2021), e.g. behavioural data, prospective adjustments | |
| | | | *Measure the ongoing progress* (Dechsling et al. 2021): Collect data from multiple measures | |
| | | Qualitative Evaluation | *Conduct behavioural observations*: film sessions (Pares et al., 2005) and take notes (Brown et al. 2016) | *Conduct behavioural observations (n=2)*: film sessions (n=2), take notes (n=1) |
| | | | *Conduct interviews*: with caregivers/parents (Pares et al. 2005), children –use video, photographs, drawings if non-verbal and contextual, closed questions, screenshots, smileys if verbal(Spiel et al. 2017) | *Carry out interviews (n=1)*: if children are able to answer (n=1) |
| **Design process** | **Participative design** | Participative design | *Use participative design* (Bozgeyikli et al. 2018; Brosnan et al. 2019; Dechsling et al. 2021; Parsons et al. 2019; Spiel et al. 2019): Value the child's design experience (Parsons et al. 2019) | *Use participative design (n=5)* |
| | | | *Explicit all questions between stakeholders before to engage in collaboration* (Parsons et al. 2019) | |
| | | Inclusivity | *Consider the entire autism spectrum*: Consider autism strengths and difficulties, include individuals requiring substantial support and adults (Bozgeyikli et al. 2018; Parsons et al. 2019) | *Consider the entire autism spectrum (n=7)*: Include individuals requiring substantial support (n=5) and adults (n=2) |
| | **Equipment** | Context-Dependent | *Use ergonomic and affordable equipment*: light (Bozgeyikli et al. 2018); portable and small (Newbutt et al. 2016); non-tethered (Bozgeyikli et al. 2018; Dautenhahn 2000), affordable (Parsons et al. 2019) | *Use ergonomic and affordable equipment (n=4):* resistive (n=3), not cumbersome (n=2), portable (n=2), affordable by using HMDs (n=2), non-tethered or with long wires (n=1) |
| | | Task-dependent | *Use AR to aid in the generalization from virtual to the real world* (Dechsling et al. 2021) | |
| | **Use context** | Setting | *Conduct experiments in ecological settings* (Bozgeyikli et al. 2018; Parsons et al. 2019) | *Conduct experiments in ecological settings (n=1)* |
| | | Age range | *Age range*: Possible from 13 years old for neurotypical people (Gent 2016) | *Age range*: is task-dependent (n=7) |
| **Information presentation** | **Little & Clear Information** | Little information | *Use little tasks to complete*: unique goal per gaming session (Carlier et al. 2020) | *Use little tasks to complete (n=4)* |
| | | | *Avoid stimuli when not task-relevant* (Bozgeyikli et al. 2018; Carlier et al. 2020; Virole 2014): use simplified graphics (Bozgeyikli et al. 2018; Tarantino et al. 2019) | *Avoid stimuli when not task-relevant*: Use neutral environment (n=7), Use simplified graphics (n=4) |
| | | | *Show clear information* (Bozgeyikli et al. 2018), e.g. foreground/background differentiation, clutter-free | *Make information clearly visible (n=1)*: Use clear foreground/background differentiation |
| | | | *Minimize transitions*: between game states (no sound, animations) (Carlier et al. 2020) | *Minimize transitions (n=1):* progressively decrease appealing stimuli before to remove the HMD |
| | | | *Avoid using metaphors* (Bozgeyikli et al. 2018) | *Avoid using metaphors (n=4)* |
| | | Adaptation to the child's pace | *Make it possible to repeat or adjust information speed* (Tardif et al. 2017) | *Make it possible to repeat or adjust information speed (n=13)* |
| | | | *Use minimal prosody*: no emotions (Duris and Clément 2018), little text/language (Carlier et al. 2020) | |
| | **Task-dependent** | Socio-emotional abilities | *Allow for rapid shifts between XR environments* (Dechsling et al. 2021) | *Use adjustable realistic naturalistic settings (n=5)*: Allows to enhance the generalization of skills |
| | | | *Use adjustable realistic naturalistic settings* (Dechsling et al. 2021): Support the adult-child relationship, give precise control over the virtual surroundings | |
| | | | *Train a variety of skills instead of specific skills* (Dechsling et al. 2021); | |
| | | | *Train balanced turn-taking*: Use various contexts, visual prompting (e.g. timer), rewards | |
| | | Others | | *Realistic and non-realistic environments won't interest the same practitioners* (n=1) |
| | **Avatars** | Representation | *Use cartoonish non-human avatars* (Bozgeyikli et al. 2018; Newbutt 2013) | *Use cartoonish non-human avatars* (n=1) |
| | | | *Make XR avatars customizable* (Bozgeyikli et al. 2018; Newbutt 2013) | |
| | | | *Only hear a XR avatar instead of both hearing and seeing it* (Newbutt 2013) | |
| | | | *Position the avatars at real-world height* (Bozgeyikli et al. 2018) | |
| | | Use | *Use avatars as tutors for children* (Bozgeyikli et al. 2018) | |